\documentclass[12pt,notitlepage,fleqn]{article}%
\usepackage{amssymb}
\usepackage{graphicx}
\usepackage{amsmath}
\usepackage{indentfirst}%
\usepackage{amsfonts}
\newtheorem{problem}{Hypothesis}

\begin{document}

\begin{center}
\ \ \ \ \ \ \ \ \ {\Huge A Proposal About the Rest}

{\Huge \ Masses of Quarks }

\bigskip

{\normalsize Jiao Lin Xu}

{\small The Center for Simulational Physics, The Department of Physics and
Astronomy}

{\small University of Georgia, Athens, GA 30602, USA}

E- mail: {\small \ Jxu@Hal.Physast.uga.edu}

\bigskip

\textbf{Abstract}
\end{center}

{\small From the Dirac sea concept, we infer that a body center cubic quark
lattice exists in the vacuum. Adapting the electron Dirac equation, we get a
special quark Dirac equation. Using its low-energy approximation, we deduced
the rest masses of the quarks: m(u)=930 Mev, m(d)=930 Mev, m(s)=1110 Mev,
m(c)=2270 Mev and m(b)=5530 Mev. We predict new excited quarks d}$_{S}%
${\small (1390), u}$_{C}${\small (6490) and d}$_{b}${\small (9950). }

\section{Introduction}

The Standard Model \cite{Standard} has been enormously successful in
explaining and predicting a wide range of phenomena. In spite of the
successes, the origin of quark masses is unknown \cite{Fayyazuddin}. In order
to answer this fundamental question,\ since quarks are born from the
vacuum,\ we will study the vacuum material. In a sense, the vacuum material
works like a superconductor. Because the transition temperature is very high
(much higher than the temperature at the center of the sun), there are no
electric or mechanical resistances to any particle or to any physical body
moving inside the vacuum material since they are moving under the transition
temperature. They moving inside it look as if they are moving in completely
empty space. The vacuum material is a super superconductor. From the Dirac sea
concept \cite{DiracSea}, we infer (see Appendix) that there is a body center
cubic (BCC) quark lattice \cite{BCC} in the vacuum. The quark lattice will
help us to deduce the rest masses of the quarks.

The purpose of this proposal is to deduce the rest masses of the quarks. We do
not discuss scattering and electroweak interactions. We will mainly discuss
the low-energy strong interactions.

\section{Fundamental Hypotheses}

\begin{problem}
: There are only two kinds of elementary quarks, u(0) and d(0), in the vacuum
state. There are super-strong attractive interactions among the quarks
(colors). These forces make and hold an infinite body center cubic (BCC) quark
lattice with a periodic constant a $\leq$ 10$^{-18}m$ in the vacuum.
\end{problem}

\begin{problem}
: Due to the effect of the vacuum quark lattice, fluctuations of energy
($\varepsilon$){\LARGE \ }and intrinsic quantum numbers (such as the Strange
number $S$) of an excited quark will exist. The fluctuation of the Strange
number is always $\Delta$S = $\pm$1 \cite{RealS}%
.\ \ \ \ \ \ \ \ \ \ \ \ \ \ \ \ \ \ \ \ \ \ \ \ \ \ \ \ \ \ \ \ \ \ \ \ \ \ \ \ \ \ \ \ \ \ \ \ \ \ \ \ \ \ \ \ \ \ \ \ \ \ \ \ \ \ \ \ \ \ \ \ \ \ \ \ \ \ \ \ \ \ \ \ \ \ \ \ \ \ \ \ \ \ \ \ \ \ \ \ \ \ \ \ \ \ \ \ \ \ \ \ \ \ \ \ \ \ \ \ \ \ \ \ \ \ \ \ \ \ \ \ \ \ \ \ \ \ \ \ \ \ \ \ \ \ \ \ \ \ \ \ \ \ \ \ \ \ \ \ \ \ \ \ \ \ \ \ \ \ \ \ \ \ \ \ \ \ \ \ \ \ \ \ \ \ \ \ \ \ \ \ \ \ \ \ \ \ \ \ \ \ \ \ \ \ \ \ \ \ \ \ \ \ \ \ \ \ \ \ \ \ \ \ \ \ \ \ \ \ \ \ \ \ \ \ \ \ \ \ \ \ \ \ \ \ \ \ \ \ \ \ \ \ \ \ \ \ \ \ \ \ \ \ \ \ \ \ \ \ \ \ \ \ \ \ \ \ \ \ \ \ \ \ \ \ \ \ \ \ \ \ \ \ \ \ \ \ \ \ \ \ \ \ \ \ \ \ \ \ \ \ \ \ \ \ \ \ \ \ \ \ \ \ \ \ \ \ \ \ \ \ \ \ \ \ \ \ \ \ \ \ \ \ \ \ \ \ \ \ \ \ \ \ \ \ \ \ \ \ \ \ \ \ \ \ \ \ \ \ \ \ \ \ \ \ \ \ \ \ \ \ \ \ \ \ \ \ \ \ \ \ \ \ \ \ \ \ \ \ \ \ \ \ \ \ \ \ \ \ \ \ \ \ \ \ \ \ \ \ \ \ \ \ \ \ \ \ \ \ \ \ \ \ \ \ \ \ \ \ \ \ \ \ \ \ \ \ \ \ \ \ \ \ \ \ \ \ \ \ \ \ \ \ \ \ \ \ \ \ \ \ \ \ \ \ \ \ \ \ \ \ \ \ \ \ \ \ \ \ \ \ \ \ \ \ \ \ \ \ \ \ \ \ \ \ \ \ \ \ \ \ \ \ \ \ \ \ \ \ \ \ \ \ \ \ \ \ \ \ \ \ \ \ \ \ \ \ \ \ \ \ \ \ \ \ \ \ \ \ \ \ \ \ \ \ \ \ \ \ \ \ \ \ \ \ \ \ \ \ \ \ \ \ \ \ \ \ \ \ \ \ \ \ \ \ \ \ \ \ \ \ \ \ \ \ \ \ \ \ \ \ \ \ \ \ \ \ \ \ \ \ \ \ \ \ \ \ \ \ \ \ \ \ \ \ \ \ \ \ \ \ \ \ \ \ \ \ \ \ \ \ \ \ \ \ \ \ \ \ \ \ \ \ \ \ \ \ \ \ \ \ \ \ \ \ \ \ \ \ \ \ \ \ \ \ \ \ \ \ \ \ \ \ \ \ \ \ \ \ \ \ \ \ \ \ \ \ \ \ \ \ \ \ \ \ \ \ \ \ \ \ \ \ \ \ \ \ \ \ \ \ \ \ \ \ \ \ \ \ \ \ \ \ \ \ \ \ \ \ \ \ \ \ \ \ \ \ \ \ \ \ \ \ \ \ \ \ \ \ \ \ \ \ \ \ \ \ \ \ \ \ \ \ \ \ \ \ \ \ \ \ \ \ \ \ \ \ \ \ \ \ \ \ \ \ \ \ \ \ \ \ \ \ \ \ \ \ \ \ \ \ \ \ \ \ \ \ \ \ \ \ \ \ \ \ \ \ \ \ \ \ \ \ \ \ \ \ \ \ \ \ \ \ \ \ \ \ \ \ \ \ \ \ \ \ \ \ \ \ \ \ \ \ \ \ \ \ \ \ \ \ \ \ \ \ \ \ \ \ \ \ \ \ \ \ \ \ \ \ \ \ \ \ \ \ \ \ \ \ \ \ \ \ \ \ \ \ \ \ \ \ \ \ \ \ \ \ \ \ \ \ \ \ \ \ \ \ \ \ \ \ \ \ \ \ \ \ \ \ \ \ \ \ \ \ \ \ \ \ \ \ \ \ \ \ \ \ \ \ \ \ \ \ \ \ \ \ \ \ \ \ \ \ \ \ \ \ \ \ \ \ \ \ \ \ \ \ \ \ \ \ \ \ \ \ \ \ \ \ \ \ \ \ \ \ \ \ \ \ \ \ \ \ \ \ \ \ \ \ \ \ \ \ \ \ \ \ \ \ \ \ \ \ \ \ \ \ \ \ \ \ \ \ \ \ \ \ \ \ \ \ \ \ \ \ \ \ \ \ \ \ \ \ \ \ \ \ \ \ \ \ \ \ \ \ \ \ \ \ \ \ \ \ \ \ \ \ \ \ \ \ \ \ \ \ \ \ \ \ \ \ \ \ \ \ \ \ \ \ \ \ \ \ \ \ \ \ \ \ \ \ \ \ \ \ \ \ \ \ \ \ \ \ \ \ \ \ \ \ \ \ \ \ \ \ \ \ \ \ \ \ \ \ \ \ \ \ \ \ \ \ \ \ \ \ \ \ \ \ \ \ \ \ \ \ \ \ \ \ \ \ \ \ \ \ \ \ \ \ \ \ \ \ \ \ \ \ \ \ \ \ \ \ \ \ \ \ \ \ \ \ \ \ \ \ \ \ \ \ \ \ \ \ \ \ \ \ \ \ \ \ \ \ \ \ \ \ \ \ \ \ \ \ \ \ \ \ \ \ \ \ \ \ \ \ \ \ \ \ \ \ \ \ \ \ \ \ \ \ \ \ \ \ \ \ \ \ \ \ \ \ \ \ \ \ \ \ \ \ \ \ \ \ \ \ \ \ \ \ \ \ \ \ \ \ \ \ \ \ \ \ \ \ \ \ \ \ \ \ \ \ \ \ \ \ \ \ \ \ \ \ \ \ \ \ \ \ \ \ \ \ \ \ \ \ \ \ \ \ \ \ \ \ \ \ \ \ \ \ \ \ \ \ \ \ \ \ \ \ \ \ \ \ \ \ \ \ \ \ \ \ \ \ \ \ \ \ \ \ \ \ \ \ \ \ \ \ \ \ \ \ \ \ \ \ \ \ \ \ \ \ \ \ \ \ \ \ \ \ \ \ \ \ \ \ \ \ \ \ \ \ \ \ \ \ \ \ \ \ \ \ \ \ \ \ \ \ \ \ \ \ \ \ \ \ \ \ \ \ \ \ \ \ \ \ \ \ \ \ \ \ \ \ \ \ \ \ \ \ \ \ \ \ \ \ \ \ \ \ \ \ \ \ \ \ \ \ \ \ \ \ \ \ \ \ \ \ \ \ \ \ \ \ \ \ \ \ \ \ \ \ \ \ \ \ \ \ \ \ \ \ \ \ \ \ \ \ \ \ \ \ \ \ \ \ \ \ \ \ \ \ \ \ \ \ \ \ \ \ \ \ \ \ \ \ \ \ \ \ \ \ \ \ \ \ \ \ \ \ \ \ \ \ \ \ \ \ \ \ \ \ \ \ \ \ \ \ \ \ \ \ \ \ \ \ \ \ \ \ \ \ \ \ \ \ \ \ \ \ \ \ \ \ \ \ \ \ \ \ \ \ \ \ \ \ \ \ \ \ \ \ \ \ \ \ \ \ \ \ \ \ \ \ \ \ \ \ \ \ \ \ \ \ \ \ \ \ \ \ \ \ \ \ \ \ \ \ \ \ \ \ \ \ \ \ \ \ \ \ \ \ \ \ \ \ \ \ \ \ \ \ \ \ \ \ \ \ \ \ \ \ \ \ \ \ \ \ \ \ \ \ \ \ \ \ \ \ \ \ \ \ \ \ \ \ \ \ \ \ \ \ \ \ \ \ \ \ \ \ \ \ \ \ \ \ \ \ \ \ \ \ \ \ \ \ \ \ \ \ \ \ \ \ \ \ \ \ \ \ \ \ \ \ \ \ \ \ \ \ \ \ \ \ \ \ \ \ \ \ \ \ \ \ \ \ \ \ \ \ \ \ \ \ \ \ \ \ \ \ \ \ \ \ \ \ \ \ \ \ \ \ \ \ \ \ \ \ \ \ \ \ \ \ \ \ \ \ \ \ \ \ \ \ \ \ \ \ \ \ \ \ \ \ \ \ \ \ \ \ \ \ \ \ \ \ \ \ \ \ \ \ \ \ \ \ \ \ \ \ \ \ \ \ \ \ \ \ \ \ \ \ \ \ \ \ \ \ \ \ \ \ \ \ \ \ \ \ \ \ \ \ \ \ \ \ \ \ \ \ \ \ \ \ \ \ \ \ \ \ \ \ \ \ \ \ \ \ \ \ \ \ \ \ \ \ \ \ \ \ \ \ \ \ \ \ \ \ \ \ \ \ \ \ \ \ \ \ \ \ \ \ \ \ \ \ \ \ \ \ \ \ \ \ \ \ \ \ \ \ \ \ \ \ \ \ \ \ \ \ \ \ \ \ \ \ \ \ \ \ \ \ \ \ \ \ \ \ \ \ \ \ \ \ \ \ \ \ \ \ \ \ \ \ \ \ \ \ \ \ \ \ \ \ \ \ \ \ \ \ \ \ \ \ \ \ \ \ \ \ \ \ \ \ \ \ \ \ \ \ \ \ \ \ \ \ \ \ \ \ \ \ \ \ \ \ \ \ \ \ \ \ \ \ \ \ \ \ \ \ \ \ \ \ \ \ \ \ \ \ \ \ \ \ \ \ \ \ \ \ \ \ \ \ \ \ \ \ \ \ \ \ \ \ \ \ \ \ \ \ \ \ \ \ \ \ \ \ \ \ \ \ \ \ \ \ \ \ \ \ \ \ \ \ \ \ \ \ \ \ \ \ \ \ \ \ \ \ \ \ \ \ \ \ \ \ \ \ \ \ \ \ \ \ \ \ \ \ \ \ \ \ \ \ \ \ \ \ \ \ \ \ \ \ \ \ \ \ \ \ \ \ \ \ \ \ \ \ \ \ \ \ \ \ \ \ \ \ \ \ \ \ \ \ \ \ \ \ \ \ \ \ \ \ \ \ \ \ \ \ \ \ \ \ \ \ \ \ \ \ \ \ \ \ \ \ \ \ \ \ \ \ \ \ \ \ \ \ \ \ \ \ \ \ \ \ \ \ \ \ \ \ \ \ \ \ \ \ \ \ \ \ \ \ \ \ \ \ \ \ \ \ \ \ \ \ \ \ \ \ \ \ \ \ \ \ \ \ \ \ \ \ \ \ \ \ \ \ \ \ \ \ \ \ \ \ \ \ \ \ \ \ \ \ \ \ \ \ \ \ \ \ \ \ \ \ \ \ \ \ \ \ \ \ \ \ \ \ \ \ \ \ \ \ \ \ \ \ \ \ \ \ \ \ \ \ \ \ \ \ \ \ \ \ \ \ \ \ \ \ \ \ \ \ \ \ \ \ \ \ \ \ \ \ \ \ \ \ \ \ \ \ \ \ \ \ \ \ \ \ \ \ \ \ \ \ \ \ \ \ \ \ \ \ \ \ \ \ \ \ \ \ \ \ \ \ \ \ \ \ \ \ \ \ \ \ \ \ \ \ \ \ \ \ \ \ \ \ \ \ \ \ \ \ \ \ \ \ \ \ \ \ \ \ \ \ \ \ \ \ \ \ \ \ \ \ \ \ \ \ \ \ \ \ \ \ \ \ \ \ \ \ \ \ \ \ \ \ \ \ \ \ \ \ \ \ \ \ \ \ \ \ \ \ \ \ \ \ \ \ \ \ \ \ \ \ \ \ \ \ \ \ \ \ \ \ \ \ \ \ \ \ \qquad

\end{problem}

\section{The Special Quark Dirac Equation}

According to the Fundamental \textbf{Hypothesis I}, there is a body center
cubic quark lattice in the vacuum. When an excited quark (q) is moving in the
vacuum, it is moving, in fact, inside the vacuum quark lattice. Physicists
usually discuss Fermion problems based on the Dirac equation. The free
particle Dirac equation \cite{FreeDirac} is
\begin{equation}
\text{i}\hbar\frac{\partial}{\partial t}\psi\text{(}\overrightarrow
{r}\text{,t) }\text{=}(\text{i}\hbar\text{c}\overrightarrow{\alpha}%
\cdot\overrightarrow{\nabla}\text{-}\beta\text{mc}^{2}\text{)}\psi
\text{(}\overrightarrow{r}\text{,t),}\label{D-Free}%
\end{equation}
where $\alpha$ are the $\alpha$-matrices, $\beta$ is the $\beta$-matrix and m
is the rest mass of the free particle in the physical vacuum. Since this
equation cannot discuss the effects of the vacuum quark lattice, we must look
for a new wave equation. We adapt the free electron Dirac equation
(\ref{D-Free}) into a quark Dirac equation that can deal with the strong
interactions between the excited quark(q$)$ and the vacuum quark lattice based
on the pure vacuum. First, we add two parts of Hamiltonian, H$_{Latt}$ and
H$_{Accom\text{ }}$. H$_{Latt}$ is the strong interactions (with body center
cubic periodic symmetries) between the excited quark (q) and the vacuum quark
lattice, and H$_{Accom\text{ }}$ is the strong interactions between the
excited quark (q) and the two accompanying excited quarks (q'$_{1}$ and
q'$_{2}$) \cite{Confine}. Then we have to change the rest mass, m, of an
electron in the physical vacuum into the bare mass \cite{BareMass}, m$_{q}$,
of an elementary quark in the pure vacuum. The quark Dirac equation will be\ \ \ \ \ \ \ \ \ \ \ \ \ \ \ \ \ \ \ \ \ \ \ \ \ %

\begin{equation}
i\hbar\frac{\partial}{\partial t}\psi\text{(}\overrightarrow{r}\text{,t) =
(H}_{0}\text{+H}_{Latt}\text{+ H}_{Accom\text{ }}\text{)}\psi\text{(}%
\overrightarrow{r}\text{,t)}\label{Q-Dirac}%
\end{equation}
where H$_{0}$ $($i$\hbar$c$\overrightarrow{\alpha}\cdot\overrightarrow{\nabla
}$-$\beta$m$_{q}$c$^{2}$) is the free Hamiltonian of an elementary quark in
the pure vacuum and m$_{q}$ is the bare mass of the elementary quarks.

Since this is a multi-particle problem, it cannot be solved exactly. We take a
mean-field approximation:
\begin{equation}
\text{H}_{Latt}\text{+ H}_{Accom\text{ }}\text{ }\simeq\text{ V}%
_{0},\label{Hs=Vo}%
\end{equation}
where V$_{0}$ is a constant at any time-space point and in any reference
frame; and
\begin{equation}%
\begin{tabular}
[c]{|l|}\hline
$\text{ }\psi\text{(}\overrightarrow{r}\text{,t) satisfies BCC symmetries.}%
$\\\hline
\end{tabular}
\label{BCC}%
\end{equation}
The above Dirac equation (\ref{Q-Dirac}) will be approximated into
\begin{equation}
\text{(i}\hbar\frac{\partial}{\partial t}\text{+ i}\hbar\text{c}%
\overrightarrow{\alpha}\cdot\overrightarrow{\nabla}\text{-}\beta\text{(m}%
_{q}\text{c}^{2}\text{+V}_{0}\text{)}\psi\text{(}\overrightarrow{r}\text{,t) =
0.}\label{SQD}%
\end{equation}
Since $($i$\hbar\frac{\partial}{\partial t}$+i$\hbar$c$\overrightarrow{\alpha
}\cdot\overrightarrow{\nabla}$-$\beta$m$_{q}$c$^{2}$)$\psi$($\overrightarrow
{r}$,t) = 0 is a free particle Dirac equation and V$_{0}$ is only a positive
constant, the above quark equation (\ref{SQD}) is Lorentz-invariant. We can
prove, as with \cite{Bjorken}, that the equation (\ref{SQD}) is
Lorentz-invariant. We call it \textbf{the special quark Dirac equation}.

Using its low-energy approximation, we can deduce the rest masses of the quarks.

\section{The Quarks}

Our purpose is to deduce the rest masses of the quarks. Since the rest mass
and the intrinsic quantum numbers of the quarks are the same in different
reference frames, we will deduce them using the classic limit. It is
sufficient to use the classic limit of the special quark Dirac equation-- the
quark Schr\"{o}dinger equation \cite{Schrodinger},%

\begin{equation}
\frac{\hslash^{2}}{\text{2m}_{q}}\nabla^{2}\Psi\text{ + (}\varepsilon
\text{-V}_{0}\text{)}\Psi\text{ = 0,}\label{Schrod}%
\end{equation}
where m$_{q}$ is the bare mass \cite{BareMass} of the elementary quark and
V$_{0}$ is a constant. Using the quark Schr\"{o}dinger equation we can deduce
the rest masses of the quarks. These are needed by particle physics
\cite{HandIn}.

\subsection{The Unflavored Quarks (u and d)}

If there were no BCC periodic symmetries (\ref{BCC}), the solution of the
above equation (\ref{Schrod}) is a free particle plane wave function of the
quarks u or d:
\begin{equation}
\varepsilon_{\overrightarrow{k}}\text{ = }\frac{\hslash^{2}}{2m_{q}}%
\text{(k}_{x}^{2}\text{+k}_{y}^{2}\text{+k}_{z}^{2}\text{)+V}_{0}\text{,
\ \ \ }\Psi_{k}\text{ = exp(i}\overrightarrow{\text{k}}\cdot\overrightarrow
{r}\text{).}\label{F-Wave}%
\end{equation}
If u is a free excited (from the vacuum) particle of the elementary
u(0)-quark, it has the same intrinsic quantum numbers that the u(0)-quark has:
\
\begin{equation}
\text{I = I}_{Z}\text{ = }\frac{1}{2}\text{, S = C = b = 0, Q = }\frac{2}%
{3}\text{.}\label{u-Q-N}%
\end{equation}
If d is a free excited (from the vacuum) particle of the elementary
d(0)-quark, it has the same intrinsic quantum numbers that the d(0)-quark has:
\
\begin{equation}
\text{I = -I}_{Z}\text{ = }\frac{1}{2}\text{, S = C = b = 0, Q = -}\frac{1}%
{3}\text{.}\label{d-Q-N}%
\end{equation}
When $\overrightarrow{k}$ = 0, $\varepsilon_{0}$ = V$_{0}$ = m$_{u}$ = m$_{d}%
$. From the masses of the proton (uu'd') and the neutron (du'd')
\cite{Confine},
\begin{equation}
\text{M}_{p}\sim\text{M}_{n}\sim\text{940 Mev,}\label{Mp=Mn}%
\end{equation}
and the accompanying excited quark masses (m$_{u\text{'}}$ = 3 Mev and
m$_{d\text{'}}$= 7 Mev) \cite{Confine} and \cite{QMass02}, we get
\begin{equation}
\text{m}_{u}\approx\text{ m}_{d}\text{ = V}_{0}\text{ = 930 Mev.}\label{Vo}%
\end{equation}
Now we have the rest masses of the unflavored quarks u and d. If we want to
find the rest masses of the flavored quarks (s, c and b), we have to consider
the BCC symmetries and the fluctuations of energies and strange numbers
(\textbf{Hypothesis II}).

\subsection{The Flavored Quarks s, c and b}

To deal with a particle moving in a lattice, physicists usually use energy
band theory. According to the energy band theory \cite{EnergyBand}, the
solution of the quark Schr\"{o}dinger equation, (\ref{Schrod}) satisfying the
BCC symmetries, will be the Bloch \cite{Bloch} wave function
\begin{equation}
\psi_{\overrightarrow{_{k}}}\text{ (}\overrightarrow{r})=e^{i\overrightarrow
{k}\cdot\overrightarrow{r}}\text{u(}\overrightarrow{r}\text{),}\label{Bloch}%
\end{equation}
where $\overrightarrow{_{k}}$ is a vector in the space of the reciprocal
lattice
\begin{equation}
\overrightarrow{k}\text{ = }\overrightarrow{l}\text{\textbf{B,}}%
\label{k-Vector}%
\end{equation}
\begin{equation}
\overrightarrow{l}\text{ = (\textit{l}}_{1}\text{, \textit{l}}_{2}\text{,
\textit{l}}_{3}\text{),}\label{l Values}%
\end{equation}
\textit{l}$_{1}$, \textit{l}$_{2}$ and \textit{l}$_{3}$ take all positive and
negative integral values including zero,
\begin{equation}
\text{\textbf{B = 2}}\pi\left(
\begin{array}
[c]{ccc}%
0 & \frac{1}{a} & \frac{1}{a}\\
\frac{1}{a} & 0 & \frac{1}{a}\\
\frac{1}{a} & \frac{1}{a} & 0
\end{array}
\right)  ,\label{B-Matrix}%
\end{equation}
$\overrightarrow{k}$ = $\overrightarrow{l}$ \textbf{B }denotes any reciprocal
lattice point; u($\overrightarrow{r}$) is a periodic function \cite{u(r)}
\begin{equation}
\text{u(}\overrightarrow{r}\text{+\textbf{A}}\overrightarrow{s}\text{) =
u(}\overrightarrow{r}\text{),}\label{u(r)}%
\end{equation}

\ \ \ \ \ \ \ \ \ \ \ \ \ \ \ \ \ \ \ \ \ \ \ \ \ \ \ \ \ \ \ \ \ \ \ \ \ \ \ \ \ \ \ \ \ \ \ \ \ \ \ \ \ \ \ \ \ \ \ \ \ \ \ \ \ \ \ \ \ \ \ \ \ %

\begin{equation}
\text{\textbf{A = }}\left(
\begin{array}
[c]{ccc}%
\text{-}\frac{a}{2} & \frac{a}{2} & \frac{a}{2}\\
\frac{a}{2} & \text{-}\frac{a}{2} & \frac{a}{2}\\
\frac{a}{2} & \frac{a}{2} & \text{-}\frac{a}{2}%
\end{array}
\right)  \text{.}\label{A-Matrix}%
\end{equation}
We have
\[
\text{T(\textbf{A}}\overrightarrow{S}\text{)}\psi_{\overrightarrow{_{k}}%
}\text{ (}\overrightarrow{r}\text{) = e}^{i\overrightarrow{k}\cdot
\overrightarrow{r}}\text{u(}\overrightarrow{r}\text{).}%
\]
Substituting the wave function (\ref{Bloch}) into the Schr\"{o}dinger equation
(\ref{Schrod}), we can get the equation of the periodic function
u($\overrightarrow{r}$)
\begin{equation}
\bigtriangledown^{2}\text{u(}\overrightarrow{r}\text{)+2i}\overrightarrow
{k}\cdot\bigtriangledown\text{u(}\overrightarrow{r}\text{)+}\frac
{\text{2m}_{q}}{\hslash^{2}}\text{[(}\varepsilon\text{ -V}_{0})\text{{\huge -}%
}\frac{\hslash^{2}\text{k}^{2}}{\text{2m}_{q}}\text{] u(}\overrightarrow
{r}\text{)=0.}\label{u-Eq}%
\end{equation}
Since V$_{0}$ is a constant, a periodic solution of (\ref{u-Eq}) is
\begin{equation}
\text{u(}\overrightarrow{r}\text{) = e}^{i\overrightarrow{l}\text{\textbf{B}%
}\cdot\overrightarrow{r}}\label{u-Wave}%
\end{equation}
for which the eigenvalue is
\begin{equation}
\varepsilon_{l}(\overrightarrow{k})=\text{V}_{0}+\frac{\hslash^{2}}%
{\text{2m}_{q}}\left\vert \overrightarrow{k}-\overrightarrow{l}%
\text{\textbf{B}}\right\vert ^{2}\text{.}\label{E(k)}%
\end{equation}
If \ taking
\begin{equation}
\vec{k}\text{ = (2}\pi\text{/a)(}\xi\text{, }\eta\text{, }\zeta\text{),}%
\label{K Value}%
\end{equation}
and substituting the B matrix (\ref{B-Matrix}) of the body center cubic quark
lattice into (\ref{E(k)}), we get the energy
\begin{equation}
\varepsilon\text{(}\vec{k}\text{,}\vec{n}\text{) =V}_{0}\text{+}%
\alpha\text{E(}\vec{k}\text{,}\vec{n}\text{),}\label{Energy}%
\end{equation}
\begin{equation}
\alpha\text{ = h}^{2}\text{/2m}_{q}\text{a}^{2}\text{,}\label{Alpha}%
\end{equation}
\begin{equation}
\text{E(}\vec{k}\text{,}\vec{n}\text{) = (n}_{1}\text{-}\xi\text{)}%
^{2}\text{+(n}_{2}\text{-}\eta\text{)}^{2}\text{+(n}_{3}\text{-}\zeta
\text{)}^{2}\text{.}\label{E-Fun}%
\end{equation}
The solution of Eq. (\ref{Schrod}) is a plane wave%

\begin{equation}
\psi_{\vec{k}}\text{(}\vec{r}\text{) = }e^{\text{(- i2}\pi\text{/a)[(n}%
_{1}\text{-}\xi\text{)x+(n}_{2}\text{-}\eta\text{)y+(n}_{3}\text{-}%
\zeta\text{)z]}}\text{,}\label{W}%
\end{equation}
where \textit{a} is the periodic constant of the quark lattice and n$_{1}$,
n$_{2}$ and n$_{3}$ are integers, n$_{1}$ = \textit{l}$_{2}$\textit{+l}$_{3}$,
n$_{2}$ =\textit{\ l}$_{3}$\textit{+l}$_{1}$ and n$_{3}$ =\textit{\ l}$_{1}
$\textit{+l}$_{2},$
\begin{equation}%
\begin{tabular}
[c]{l}%
\textit{l}$_{1}$ =1/2( -n$_{1}$+n$_{2}$+n$_{3}$),\\
\textit{l}$_{2}$ =1/2( +n$_{1}$-n$_{2}$+n$_{3}$),\\
\textit{l}$_{3}$ =1/2( +n$_{1}$+n$_{2}$-n$_{3}$),
\end{tabular}
\label{l-n}%
\end{equation}
satisfying the condition that only those values of $\overrightarrow{n}$ =
(n$_{1}$, n$_{2}$, n$_{3}$) are allowed, which make $\overrightarrow{l}$ =
\textit{(l}$_{1}$\textit{, l}$_{2}$\textit{, l}$_{3}$\ ) an integer vector
\cite{LValues}. Condition (\ref{l-n}) implies that the vector $\vec{n}$ =
(n$_{1}$, n$_{2}$, n$_{3}$) can only take certain values. For example $\vec
{n}$ cannot take (0, 0, 1) or (1, 1, -1), but it can take (0, 0, 2) and (1,
-1, 2).

\subsubsection{The Energy Bands \textbf{\qquad}}

Using the standard methods of energy band theory \cite{EnergyBand}, we can
deduce whole energy bands and wave functions for low energy bands, from
(\ref{Energy}) and (\ref{W}). Since the purpose of this paper is deducing the
rest masses of the quarks, we do not show the wave functions. Because quarks
are all born on the single-energy bands of the $\Delta$-axis and the $\Sigma
$-axis,\ we will discuss these single-energy bands only.\ %

\begin{equation}%
\begin{tabular}
[c]{l}%
\ \ \ \ \ \ \ \ The single energy bands on the $\Delta$-axis ($\Gamma$-H)\\
$%
\begin{tabular}
[c]{|l|l|l|l|l|l|l|}\hline
$\text{{\small Energy Bands}}$ & $\text{n}_{1}\text{,n}_{2}\text{,n}_{3}$ &
$\text{{\small R}}$ & $\text{{\small d}}$ & \ \ J$_{\Delta}$ &
$\text{{\small E}}$ & $\text{{\small Energy}}$\\\hline
$\text{E}_{\Gamma}\text{=0}\rightarrow\text{E}_{H}\text{=1}$ & 0, 0, \ 0 & 4 &
1 & \ \ 0 & 0 & 930\\\hline
$\text{E}_{H}\text{=1}\rightarrow\text{E}_{\Gamma}\text{=4}$ & 0, 0, \ 2 & 4 &
1 & J$_{H}$=1 & 1 & 1290\\\hline
$\text{E}_{\Gamma}\text{=4}\rightarrow\text{E}_{H}\text{=9}$ & 0, 0, -2 & 4 &
1 & J$_{\Gamma}$=1 & 4 & 2370\\\hline
$\text{E}_{H}\text{=9}\rightarrow\text{E}_{\Gamma}\text{=16}$ & 0, 0, \ 4 &
4 & 1 & J$_{H}$=2 & 9 & 4170\\\hline
$\text{E}_{\Gamma}\text{=16}\rightarrow\text{E}_{H}\text{=25}$ & 0, 0, -4 &
4 & 1 & J$_{\Gamma}$=2 & 16 & 6690\\\hline
$\text{E}_{H}\text{=25}\rightarrow\text{E}_{\Gamma}\text{=36}$ & 0, 0, \ 6 &
4 & 1 & J$_{H}$=3 & 25 & 9930\\\hline
... & ... & ... & ... & ... & ... & ...\\\hline
\end{tabular}
\ $%
\end{tabular}
\label{D-1}%
\end{equation}

and%

\begin{align}
& \text{ \ \ \ \ \ \ The single energy bands on the }\Sigma\text{-axis(}%
\Gamma\text{-N)}\nonumber\\
&
\begin{tabular}
[c]{|l|l|l|l|l|l|l|}\hline
$\text{{\small Energy Bands}}$ & $\text{n}_{1}\text{, n}_{2}\text{, n}_{3}$ &
R & d & $\text{\ \ J}_{\Sigma}$ & E & $\text{{\small Energy}}$\\\hline
$\text{E}_{\Gamma}\text{=0}\rightarrow\text{E}_{N}\text{=1/2}$ & 0, \ 0, \ 0 &
2 & 1 & \ \ 0 & 0 & 930\\\hline
$\text{E}_{N}\text{=1/2}\rightarrow\text{E}_{\Gamma}\text{=2}$ & 1, \ 1, \ 0 &
2 & 1 & $\text{J}_{N}\text{=1}$ & $\frac{1}{2}$ & 1110\\\hline
$\text{E}_{\Gamma}\text{=2}\rightarrow\text{E}_{N}\text{=9/2}$ & $\text{-1,
-1, 0}$ & 2 & 1 & $\text{J}_{\Gamma}\text{=1}$ & 2 & 1650\\\hline
$\text{E}_{N}\text{=9/2}\rightarrow\text{E}_{\Gamma}\text{=8}$ & 2, \ 2, \ 0 &
2 & 1 & $\text{J}_{N}\text{=2}$ & $\frac{9}{2}$ & 2550\\\hline
$\text{E}_{\Gamma}\text{=8}\rightarrow\text{E}_{N}\text{=25/2}$ & -2, -2, 0 &
2 & 1 & $\text{J}_{\Gamma}\text{=2}$ & 8 & 3810\\\hline
$\text{E}_{N}\text{=25/2}\rightarrow\text{E}_{\Gamma}\text{=18}$ & 3, \ 3,
\ 0 & 2 & 1 & $\text{J}_{N}\text{=3}$ & $\frac{25}{2}$ & 5430\\\hline
$\text{E}_{\Gamma}\text{=18}\rightarrow\text{E}_{N}\text{=49/2}$ & -3,
-3,\ 0 & 2 & 1 & $\text{J}_{\Gamma}\text{=3}$ & 18 & 7410\\\hline
$\text{E}_{N}\text{=49/2}\rightarrow\text{E}_{\Gamma}\text{=64}$ & 4, \ 4,
\ 0 & 2 & 1 & $\text{J}_{N}\text{=4}$ & $\frac{49}{2}$ & 9750\\\hline
... & ... & ... & ... & ... & ... & ...\\\hline
\end{tabular}
\label{Sigma-1}%
\end{align}

\subsubsection{The Phenomenological Formulae of the Quantum Numbers}

We have already found the energy bands that are shown in (\ref{D-1}) and
(\ref{Sigma-1}). These energy bands represent the excited states of the
elementary quarks u(0) and d(0). These energy band excited states are the
quarks and excited quarks. Comparing them with experimental data, we can
recognize the quarks. For the first Brillouin zone, $\overrightarrow{n}$ = (0,
0, 0); it is a part of the free quark solution (\ref{F-Wave}) and it has the
lowest energy (mass). It represents the lowest mass u-quark (\ref{u-Q-N}) and
the d-quark (\ref{d-Q-N}). Fitting the energy band excited states to the
experimental results, we find the $\alpha$ in (\ref{Energy})
\begin{equation}
\alpha\text{ = }\frac{\text{h}^{2}}{\text{2m}_{q}\text{a}^{2}}\text{ = 360
Mev.}\label{360 Mev}%
\end{equation}

Then we can find the formulae that we can use to deduce the quantum numbers
and the rest masses of the energy bands, as shown in the following:

1. Baryon number $B$: When an elementary quark is excited (or accompanying
excited \cite{Confine}) from the vacuum state, it has
\begin{equation}
B=\frac{1}{3}.\label{B-Number}%
\end{equation}
\ 

2. Isospin number $I$: $I$\ is determined by the energy band degeneracy
deg\ \cite{EnergyBand} with
\begin{equation}
\text{deg = 2I + 1}.\label{IsoSpin}%
\end{equation}

3. Strange number $S$: $S$\ is determined by the rotary fold $R$\ of the
symmetry axis \cite{EnergyBand} with
\begin{equation}
S=R-4,\label{SNumber}%
\end{equation}
where the number $4$\ is the highest possible rotary fold number of the BCC lattice.

4. Electric charge $Q$: The electric charge Q$_{q}$ of the excited quark q is
determined completely by the elementary quark (u(0) or d(0)) that is excited
to produce the excited quark q. After getting B, I$_{Z}$ and S (C and b), and
considering the generalized Gell-Mann-Nishigima relationship \cite{GMN}, we
can find the electric charges of the energy bands. If the I$_{z}$+$\frac{1}%
{2}$(B+S+C+b)
$>$%
0, it is an excited state of the elementary u(0)-quark,
\begin{equation}
\text{I}_{z}\text{+}\frac{1}{2}\text{(B+S+C+b)
$>$
0, \ \ \ Q}_{\text{q}}\text{= Q}_{\text{u}}=\frac{2}{3}\text{;}\label{q=u}%
\end{equation}
otherwise, it is a excited state of the elementary d(0)-quark
\begin{equation}
\text{I}_{z}\text{+}\frac{1}{2}\text{(B+S+C+b)
$>$
0, \ \ \ Q}_{\text{q}}\text{= Q}_{\text{d}}\text{= -}\frac{1}{3}%
\text{.}\label{q=d}%
\end{equation}

5. If a degeneracy (or subdegeneracy) of a group of energy bands is smaller
than the rotary fold R,
\begin{equation}
\text{deg}\ \text{%
$<$%
\ R\ \ and\ \ R- deg}\ \neq\text{ 2,}\label{Condi-Sbar}%
\end{equation}
\ then the formula (\ref{SNumber}) will be replaced by
\begin{equation}
\text{\={S} = R - 4.}\label{Strangebar}%
\end{equation}
\ The real value of $S$\ is
\begin{equation}
\text{S = \={S} + }\Delta\text{S = S}_{Axis}\pm\text{1.}\label{strangeflu}%
\end{equation}
From Hypothesis II, $\Delta S=\pm1$, we have a\textbf{\ }formula to deduce
$\Delta$S,
\begin{equation}
\Delta\text{S = [1-2}\delta\text{(S)]Sign(}\vec{n}\text{),\ \ }%
\label{DaltaS-Value}%
\end{equation}
where
\begin{equation}
\text{Sign(}\vec{n}\text{) = }\frac{\text{n}_{1}\text{+n}_{2}\text{+n}_{3}%
}{\left\vert \text{n}_{1}\right\vert \text{+}\left\vert \text{n}%
_{2}\right\vert \text{+}\left\vert \text{n}_{3}\right\vert }\text{.}%
\label{Sign(n)}%
\end{equation}

6. The fluctuation of the strange number will be accompanied by an energy
change (\textbf{Hypothesis II}). We assume that the change of the energy
(perturbation energy) is proportional to (-$\Delta$S)\ and a number,
J,\ representing the energy level, as a phenomenological formula:
\[
\Delta\varepsilon\text{ = (S+1)100(J+S)(-}\Delta\text{S).}%
\]
For a single energy band, J will take 1, 2, 3 and so forth from the lowest
energy band to higher ones for each of the two end points of the symmetry axes
respectively. \ 

7. Charmed number $C$\ and Bottom number $b$: The \textquotedblleft Strange
number,\textquotedblright\ S, in (\ref{strangeflu}) is not completely the same
as the strange number in (\ref{SNumber}). In order to compare it with the
experimental results, we would like to give it a new name under certain
circumstances.\qquad

If S = +1, which originates from the fluctuation $\Delta S=+1$,
\begin{equation}
\text{\ then we call it the Charmed number }C\text{ }(C=+1)\text{;
\ \ \ \ \ }\label{Charmed}%
\end{equation}
if S = -1$,$ which originates from the fluctuation $\Delta S=+1$, and there is
an energy fluctuation,$\ \ \ \ \ \ \ \ \ \ \ \ \ \ \ $%
\begin{equation}
\text{then we call it the Bottom number }b\text{ }(b=-1)\text{.}\label{Battom}%
\end{equation}

8. We assume that the excited quark's rest mass is the minimum energy of the
energy band that represents the quark: \ \ \ \ \ \ \ \ \ \ \ \ \ \ \ \ \ \ \ \ \ \ \ \ \ \ \ \ \ \ \ \ \ \ %

\begin{equation}
m_{q^{\ast}}=\text{Minimum[V}_{0}\text{+ }\alpha\text{E( }\overrightarrow
{k}\text{,}\overrightarrow{n}\text{)] + }\Delta\varepsilon\text{.}\label{U-M}%
\end{equation}
\textbf{This formula (\ref{U-M}) is the united mass formula} that can give the
masses of all quarks.\ \ 

\subsubsection{The Recognition of the Quarks\textbf{\qquad}%
\ \ \ \ \ \ \ \ \ \ \ \ \ \ \ \ \ \ \ }

Using the above formulae (\ref{B-Number}) - (\ref{U-M}), we can find the
quantum numbers and masses of the energy bands. Using the quantum numbers and
the masses, we can recognize the quarks (ground and excited quarks). Since the
purpose of this paper is deducing the rest masses of the quarks, we do not
show the whole quark spectrum. Because quarks are all born on the
single-energy bands of the $\Delta$-axis and the $\Sigma$-axis,\ we will
discuss the quarks on the single-energy bands only.

1. The single-bands on the $\Delta$-axis ($\Gamma$-H)\textbf{\qquad}

For the single-bands on the $\Delta$-axis, R=4, S$_{\Delta}$ = 0 from
(\ref{SNumber}); d =1, I =0 from (\ref{IsoSpin}). Since d = 1
$<$
R = 4 and R-d = 3 $\neq$ \ 2, according to (\ref{Condi-Sbar}), we will use
(\ref{strangeflu}) instead of (\ref{U-M}). Using (\ref{Charmed}), we have
\qquad\qquad%

\begin{equation}%
\begin{tabular}
[c]{|l|l|l|l|l|l|l|l|l|}\hline
Energy Band & $\text{n}_{1}\text{,n}_{2}\text{,n}_{3}$ & $\text{S}_{\Delta}$ &
$\Delta\text{S}$ & \ \ \ J & $\text{\ }\Delta\varepsilon$ & $\text{ S}$ & C &
$\ \ \text{q(m)}$\\\hline
$\text{E}_{\Gamma}\text{=0}\rightarrow\text{E}_{\text{H}}\text{=1}$ & 0, \ 0,
\ 0 & $\text{0}$ & 0 & \ \ \ 0 & \ \ \ 0 & 0 & 0 & $\ \text{u(930)}$\\\hline
$\text{E}_{\text{H}}\text{=1}\rightarrow\text{E}_{\Gamma}\text{=4}$ &
$\text{0, \ 0, \ 2}$ & 0 & -1 & $\text{J}_{\text{H}}\text{=1}$ & 100 & -1 &
0 & $\text{d}_{S}\text{(1390)}$\\\hline
$\text{E}_{\Gamma}\text{=4}\rightarrow\text{E}_{\text{H}}\text{=9}$ &
$\text{0, \ 0, -2}$ & 0 & +1 & $\text{J}_{\Gamma}\text{=1}$ & -100 & 0 & 1 &
$\text{u}_{C}\text{(2270)}$\\\hline
$\text{E}_{\text{H}}\text{=9}\rightarrow\text{E}_{\Gamma}\text{=16}$ &
$\text{0, \ 0, \ 4}$ & 0 & -1 & $\text{J}_{\text{H}}\text{=2}$ & 200 & -1 &
0 & $\text{d}_{S}\text{(4370)}$\\\hline
$\text{E}_{\Gamma}\text{=16}\rightarrow\text{E}_{\text{H}}\text{=25}$ &
$\text{0, \ 0, -4}$ & 0 & +1 & $\text{J}_{\Gamma}\text{=2}$ & -200 & 0 & 1 &
$\text{u}_{C}\text{(6490)}$\\\hline
$\text{E}_{\text{H}}\text{=25}\rightarrow\text{E}_{\Gamma}\text{=16}$ &
$\text{0, \ 0, \ 6}$ & 0 & -1 & $\text{J}_{\text{H}}\text{=3}$ & 300 & -1 &
0 & $\text{d}_{S}\text{(10230)}$\\\hline
... & ... & ... & ... & ... & ... & ... & ... & ...\\\hline
\end{tabular}
\label{D1}%
\end{equation}

2. The single bands on the axis $\Sigma$($\Gamma$-N)\textbf{\qquad}

For the single bands on the $\Sigma$-axis, R=2, S$_{\Sigma}$ = -2 from
(\ref{SNumber}); d =1, I =0 from (\ref{IsoSpin}). Since d = 1
$<$
R = 2 and R-d = 1 $\neq$ \ 2, according to (\ref{Condi-Sbar}), we should use
(\ref{strangeflu}) instead of (\ref{SNumber}). Using (\ref{Battom}), we have
\qquad\qquad%

\begin{equation}%
\begin{tabular}
[c]{|l|l|l|l|l|l|l|}\hline
Energy Band & $\text{n}_{1}\text{, n}_{2}\text{, n}_{3}$ & $\text{S}_{\Sigma}$
& $\Delta\text{S}$ & S & \ \ J & \ $\text{q(m)}$\\\hline
$\text{E}_{\Gamma}\text{=0}\rightarrow\text{E}_{N}\text{=}\frac{\text{1}}{2}$
& $\text{0, \ 0, \ 0}$ & 0 & 0 & 0 & \ \ 0 & $\text{d(930)}$\\\hline
$\text{E}_{N}\text{=}\frac{\text{1}}{2}\rightarrow\text{E}_{\Gamma}\text{=2}$
& $\text{1, \ 1, \ 0}$ & -2 & 1 & -1 & $\text{J}_{N}\text{=1}$ & $\text{d}%
_{S}\text{(1110)}$\\\hline
$\text{E}_{\Gamma}\text{=2}\rightarrow\text{E}_{N}\text{=}\frac{\text{9}}{2}$
& -1$\text{, -1, \ 0}$ & -2 & -1 & -3 & $\text{J}_{\Gamma}\text{=1}$ &
$\text{d}_{\Omega}\text{(1650)}$\\\hline
$\text{E}_{N}\text{=}\frac{\text{9}}{2}\rightarrow\text{E}_{\Gamma}\text{=8}$
& $\text{2, \ 2, \ 0}$ & -2 & 1 & -1 & $\text{J}_{N}\text{=2}$ & $\text{d}%
_{S}\text{(2550)}$\\\hline
$\text{E}_{\Gamma}\text{=8}\rightarrow\text{E}_{N}\text{=}\frac{\text{25}}{2}$
& $\text{-2, -2, \ 0}$ & -2 & -1 & -3 & $\text{J}_{\Gamma}\text{=2}$ &
$\text{d}_{\Omega}\text{(3810)}$\\\hline
$\text{E}_{N}\text{=}\frac{\text{25}}{2}\rightarrow\text{E}_{\Gamma}%
\text{=18}$ & $\text{3, \ 3, \ 0}$ & -2 & 1 & -1 & $\text{J}_{N}\text{=3}$ &
$\text{d}_{b}\text{(5530)}$\\\hline
$\text{E}_{\Gamma}\text{=18}\rightarrow\text{E}_{N}\text{=}\frac{\text{49}}%
{2}$ & $\text{-3, -3, \ 0}$ & -2 & -1 & -3 & $\text{J}_{\Gamma}\text{=3}$ &
$\text{d}_{\Omega}\text{(7310)}$\\\hline
$\text{E}_{N}\text{=}\frac{\text{49}}{2}\rightarrow\text{E}_{\Gamma}%
\text{=32}$ & $\text{ 4, \ 4, \ 0}$ & -2 & 1 & -1 & $\text{J}_{N}\text{=4}$ &
$\text{d}_{b}\text{(9950)}$\\\hline
... & ... & ... & ... & ... & ... & \ \ \ \ \ .\\\hline
\end{tabular}
\label{Sygema-1}%
\end{equation}
\ \ \ 

\subsubsection{The Quark Spectrum\textbf{\qquad}\ \ \ \ \ }

Continuing the above procedure, we can find the lower energy excited states of
the elementary quarks. From (\ref{u-Q-N}), (\ref{d-Q-N}), (\ref{Vo}),
(\ref{D1}) and (\ref{Sygema-1}), there are only five ground states: u(930),
d(930), s(1110), c(2270) and b(5530). They are the quarks; the others are the
excited states. Since the purpose of this paper is to deduce the rest masses
of the quarks (about excited quarks will show in our next paper
\textquotedblleft The Baryon Spectrum\textquotedblright), we list the quarks
as shown in Table 1:

\begin{center}
$
\begin{tabular}
[c]{|l|}\hline
\ \ \ \ \ \ \ \ \ \ Table 1. The quark Spectrum\\\hline
\ \ {\small Elementary quarks, u(0) and d(0), in the vacuum}\\\hline%
\begin{tabular}
[c]{|l|}\hline
\ \ \ \ \ \ {\small The two accompanying excited quarks}\\\hline
{\small u}', {\small S=C=b=0, }I=s=$\frac{1}{2}$, {\small I}$_{z}$=$\frac
{1}{2}$, {\small Q=}$\frac{2}{3}$, {\small m}$_{u^{\prime}}${\small =3}%
\\\hline
{\small d}', {\small S=C=b=0, }I=s=$\frac{1}{2}$, {\small I}$_{z}$=$\frac
{-1}{2}$, {\small Q=}$\frac{-1}{3}$, {\small m}$_{d^{^{\prime}}}$=7\\\hline
\end{tabular}
\\\hline
\ \
\begin{tabular}
[c]{l}%
\ \ \ \ \ \ The Quarks (The Ground States)\\%
\begin{tabular}
[c]{|l|l|l|l|l|l|}\hline
q$_{\text{uark}}$ & \ \textbf{u} & \ \textbf{d} & \textbf{\ \ s} &
\textbf{\ \ c} & $\mathbf{\ }$\textbf{b}\\\hline
\ \ m & 930 & 930 & 1110 & 2270 & 5530\\\hline
\ \ S & \ \ 0 & \ \ 0 & \ -1 & \ \ 0 & \ \ 0\\\hline
\ \ C & \ \ 0 & \ \ 0 & \ \ 0 & \ \ 1 & \ \ 0\\\hline
\ \ b & \ \ 0 & \ \ 0 & \ \ 0 & \ \ 0 & \ -1\\\hline
\ \ I & +$\frac{1}{2}$ & +$\frac{1}{2}$ & \ \ 0 & \ \ 0 & \ \ 0\\\hline
\ \ I$_{Z}$ & +$\frac{1}{2}$ & -$\frac{1}{2}$ & \ \ 0 & \ \ 0 & \ \ 0\\\hline
\ Q & +$\frac{2}{3}$ & -$\frac{1}{3}$ & -$\frac{1}{3}$ & +$\frac{2}{3}$ &
-$\frac{1}{3}$\\\hline
\end{tabular}
\end{tabular}
\\\hline
\end{tabular}
$
\end{center}

\subsection{The Baryons
\ \ \ \ \ \ \ \ \ \ \ \ \ \ \ \ \ \ \ \ \ \ \ \ \ \ \ \ \ \ \ \ \ \ \ \ \ \ \ \ \ \ \ \ \ \ \ \ \ \ \ \ \ \ \ \ \ \ \ \ \ \ \ \ \ \ \ \ \ \ \ \ \ \ \ \ \ \ \ \ \ \ \ \ \ \ \ \ \ \ \ \ \ \ \ \ \ \ \ \ \ \ \ \ \ \ \ \ \ \ \ \ \ \ \ \ \ \ \ \ \ \ \ \ \ \ \ \ \ \ \ \ \ \ \ \ \ \ \ \ \ \ \ \ \ \ \ \ \ \ \ \ \ \ \ \ \ \ \ \ \ \ \ \ \ \ \ \ \ \ \ \ \ \ \ \ \ \ \ \ \ \ \ \ \ \ \ \ \ \ \ \ \ \ \ \ \ \ \ \ \ \ \ \ \ \ \ \ \ \ \ \ \ \ \ \ \ \ \ \ \ \ \ \ \ \ \ \ \ \ \ \ \ \ \ \ \ \ \ \ \ \ \ \ \ \ \ \ \ \ \ \ \ \ \ \ \ \ \ \ \ \ \ \ \ \ \ \ \ \ \ \ \ \ \ \ \ \ \ \ \ \ \ \ \ \ \ \ \ \ \ \ \ \qquad
\ \ \ \ \ \ \ \ \ \ \ \ \ \ \ \ \ \ \ \ \ \ \ \ \ \ \ \ \ \ \ \ \ \ \ \ \ \ \ \ \ \ \ \ \ \ \ \ \ \ \ \ \ \ \ \ \ \ \ \ \ \ \ \ \ \ }%

Any excited q is always accompanied by two accompanying excited quarks
(q'$_{1}$ and q'$_{2}$) \cite{Confine}. The baryon number of the three-quark
system (qq'$_{1}$q'$_{2}$), from (\ref{B-Number}), equals the sum of the three
quarks,
\begin{equation}
\text{B}_{\text{qq}_{1}^{\text{,}}\text{q}_{2}^{\text{,}}}\text{ = B}_{\text{q
}}\text{+ B}_{\text{q}_{1}^{\text{,}}}\text{+ B}_{\text{q}_{2}^{\text{,}}%
}\text{ = 1.}\label{3-q}%
\end{equation}
(\ref{3-q}) means that the three-quark systems are the baryons. We give the
most important baryons in Table 2:

\ \ \ $\
\begin{tabular}
[c]{l}%
$\ \ \ \ \ \ \ \text{Table 2.\ The Ground States of the Baryons}$\\
$%
\begin{tabular}
[c]{|l|l|l|l|}\hline
{\small \ \ \ \ \ Theory Baryons} & {\small \ Quantum. No} &
{\small Experiment} & $\Delta M$\\\hline
{\small Baryon(M) [q(m)q'}$_{1}${\small q'}$_{2}${\small ]} & {\small \ S,
\ C,\ \ b, \ I, \ Q} & {\small Name(M)} & {\small R}\\\hline
{\small N}$^{+}${\small (940) [u}$_{N}${\small (930) u(3)}$^{\text{,}}%
${\small d(7)}$^{\text{,}}${\small ]} & {\small \ 0,\ \ 0, \ 0, }$\frac{1}{2}%
${\small , 1} & {\small p(938)} & {\small 0.2}\\\hline
{\small N}$^{0}${\small (940) [d}$_{N}${\small (930)u(3)}$^{\text{,}}%
${\small d(7)}$^{\text{,}}${\small ]} & {\small 0, \ 0, \ 0, }$\frac{1}{2}%
${\small , 0} & {\small n(940)} & {\small 0.0}\\\hline
$\Lambda_{s}^{0}${\small (1120) [d}$_{\Lambda}${\small (1110)u(3)}$^{\text{,}%
}${\small d(7)}$^{\text{,}}${\small ]} & {\small -1,\ \ 0, \ 0, \ 0, \ 0} &
$\Lambda^{0}${\small (1116)} & {\small 0.4}\\\hline
$\Lambda_{c}^{+}${\small (2280) [u}$_{C}${\small (2270)u(3)}$^{\text{,}}%
${\small d(7)}$^{\text{,}}${\small ]} & {\small 0, \ 1, \ 0, \ 0, \ 1} &
$\Lambda_{c}^{+}${\small (2285)} & {\small 0.2}\\\hline
$\Lambda_{b}^{0}${\small (5540) [d}$_{b}${\small (5530)u(3)}$^{\text{,}}%
${\small d(7)}$^{\text{,}}${\small ]} & {\small 0, \ 0, -1, \ 0, \ 0} &
$\Lambda_{b}^{0}${\small (5641)} & {\small 1.8}\\\hline
\end{tabular}
\ \ \ \ $\\
\ \ \ \ \ {\small In the fourth column, R =(}$\frac{\Delta\text{M}}{\text{M}}%
${\small )\%.}%
\end{tabular}
\ $\ \ \ \ \ 

The theoretical quantum numbers (I, S, C, B, Q) of the important baryons are
completely the same with the experimental results. The theoretical masses of
the important baryons agree\ well with the experimental results. These show
that the theoretical masses of the quarks\ are
correct.\ \ \ \ \ \ \ \ \ \ \ \ \ \ \ \ \ \ \ \ \ \ \ \ \ \ \ \ \ \ \ \ \ \ \ \ \ \ \ \ \ \ \ \ \ \ \ \ \ \ \ \ \ \ \ \ \ \ \ \ \ \ \ \ \ \ \ \ \qquad
\qquad\qquad\qquad
\ \ \ \ \ \ \ \ \ \ \ \ \ \ \ \ \ \ \ \ \ \ \ \ \ \ \ \ \ \ \ \ \ \ \ \ \ \ \ \ \ \ \ \ \ \ \ \ \ \ \ \ \ \ \ \ \ \ \ \ \ \ \ \ \ \ \ \ \ \ \ \ \ \ \ \ \ \ \ \ \ \ \ \ \ \ \ \ \ \ \ \ \ \ \ \ \ \ \ \ \ \ \ \ \ \ \ \ \ \ \ \ \ \ \ \ \ \ \ \ \ \ \ \ \ \ \ \ \ \ \ \ \ \ \ \ \ \ \ \ \ \ \ \ \ \ \ \ \ \ \ \ \ \ \ \ \ \ \ \ \ \ \ \ \ \ \ \ \ \ \ \ \ \ \ \ \ \ \ \ \ \ \ \ \ \ \ \ \ \ \ \ \ \ \ \ \ \ \ \ \ \ \ \ \ \ \ \ \ \ \ \ \ \ \ \ \ \ \ \ \ \ \ \ \ \ \ \ \ \ \ \ \ \ \ \ \ \ \ \ \ \ \ \ \ \ \ \ \ \ \ \qquad
\ \ \ \ \ \ \ \ \ \ \ \ \ \ \ \ \ \ \ \ \ \ \ \ \ \ \ \ \ \ \ \ \ \ \ \ \ \ \ \ \ \ \ \ \ \ \ \ \ \ \ \ \ \ \ \ \ \ \ \ \ \ \ \ \ \ 

\subsection{The Mesons
\ \ \ \ \ \ \ \ \ \ \ \ \ \ \ \ \ \ \ \ \ \ \ \ \ \ \ \ \ \ \ \ \ \ \ \ \ \ \ \ \ \ \ \ \ \ \ \ \ \ \ \ \ \ \ \ \ \ \ \ \ \ \ \ \ \ \ \ \ \ \ \ \ \ \ \ \ \ \ \ \ \ \ \ \ \ \ \ \ \ \ \ \ \ \ \ \ \ \ \ \ \ \ \ \ \ \ \ \ \ \ \ \ \ \ \ \ \ \ \ \ \ \ \ \ \ \ \ \ \ \ \ \ \ \ \ \ \ \ \ \ \ \ \ \ \ \ \ \ \ \ \ \ \ \ \ \ \ \ \ \ \ \ \ \ \ \ \ \ \ \ \ \ \ \ \ \ \ \ \ \ \ \ \ \ \ \ \ \ \ \ \ \ \ \ \ \ \ \ \ \ \ \ \ \ \ \ \ \ \ \ \ \ \ \ \ \ \ \ \ \ \ \ \ \ \ \ \ \ \ \ \ \ \ \ \ \ \ \ \ \ \ \ \ \ \ \ \ \ \ \ \ \ \ \ \ \ \ \ \ \ \ \ \ \ \ \ \ \ \ \ \ \ \ \ \ \ \ \ \ \ \ \ \ \ \ \ \ \ \ \ \ \ \ \ \ \qquad
\ \ \ \ \ \ \ \ \ \ \ \ \ \ \ \ \ \ \ \ \ \ \ \ \ \ \ \ \ \ \ \ \ \ \ \ \ \ \ \ \ \ \ \ \ \ \ \ \ \ \ \ \ \ \ \ \ \ \ \ \ \ \ \ \ \ }%

\ According to the Quark Model, a meson is made of a quark$\ $q$_{i}$ and an
antiquark $\overline{q_{j}}$. Since we have found the quark spectrum (see
Table 1), using the sum laws, we can find the quantum numbers (S, C, b, I and
Q) of the quark pairs (q$_{i}$q$_{j}$). Since there is not a theoretical
formula for the binding energies, we propose a phenomenological formula for
the binding energy. Because all quarks are the excited states of the
elementary quarks u(0) and d(0), the binding energies are roughly constant (-
1671 Mev). If the differences between the quark mass and the antiquark mass in
the quark pairs is larger, the binding energy is smaller ($\frac{\Delta
m}{930}$):
\begin{equation}
\text{E}_{B}\text{(q}_{i}\overline{q_{j}}\text{)= -1671+100[}\frac
{\Delta\text{m}}{\text{930}}\text{+}\Delta\text{G+(C}_{i}\text{+}%
\overline{\text{C}_{j}}\text{)]+50}\delta\text{(G}_{i}\text{)}\delta
\text{(G}_{j}\text{),\ }\label{B-E}%
\end{equation}
\ where $\Delta$m = $\left\vert \text{m}_{i}\text{- m}_{j}\right\vert ;$
$\Delta$G = $\left\vert \text{G}_{i}\text{+G}_{j}\right\vert $\ $\left\vert
\text{G}_{i}\text{+}\overline{G_{j}}\right\vert $, G = S + C\ +b, S-- strange
number, C-- charmed number and b-- bottom number; $\delta$(G$_{i}$) is Dirac
function if G$_{i}$ = 0 $\delta$(G$_{i}$) =1 and if G$_{i}$ $\neq$ 0 $\delta
$(G$_{i}$) =0. Using (\ref{B-E}) we can deduce the masses of the most
important mesons as shown in Table 3:

\ \ \ \ \ \ \ \ \ 

$
\begin{tabular}
[c]{l}%
$\ \ \ \ \ \ \ \ \ \ \ \ \ \ \ \ \ \text{Table 3.\ \ The Most Important
Mesons}$\\
$%
\begin{tabular}
[c]{|l|l|l|l|l|}\hline
Experiment & $\ \ \ \overline{\text{q}_{j}\text{(m}_{j}\text{)}}\text{q}%
_{i}\text{(m}_{i}\text{)}$ & E$_{bind}$ & Theory & \ R\\\hline
$\pi$(139) & $\overline{\text{q}_{N}\text{(930)}}\text{q}_{N}^{{}}%
\text{(930)}$ & -1721 & $\pi$(139) & \ 0\\\hline
K(494) & $\overline{\text{q}_{S}\text{(1110)}}\text{q}_{N}\text{(930)}$ &
-1552 & K(488) & 1.2\\\hline
$\eta$(547) & $\overline{\text{q}_{S}\text{(1110)}}\text{q}_{S}\text{(1110)} $
& -1671 & $\eta$(549) & 0.4\\\hline
D(1869) & $\overline{\text{q}_{N}\text{(930)}}\text{q}_{C}\text{(2270)}$ &
-1327 & D(1878) & 0.2\\\hline
D$_{S}$(1969) & $\overline{\text{q}_{S}\text{(1110)}}\text{q}_{C}%
\text{(2270)}$ & -1446 & D$_{S}$(1934) & 1.8\\\hline
J/$\psi$(3097) & $\overline{\text{q}_{C}\text{(2270)}}\text{q}_{C}%
\text{(2270)}$ & -1471 & J/$\psi$(3069) & 0.9\\\hline
B(5279) & $\overline{\text{q}_{b}\text{(5530)}}\text{q}_{N}\text{(930)}$ &
-1076 & B(5384) & 2.0\\\hline
B$_{S}$(5344) & $\overline{\text{q}_{b}\text{(5530)}}\text{q}_{S}%
\text{(1110)}$ & -1196 & B$_{S}$(5444) & 1.9\\\hline
B$_{C}$(6400) & $\overline{\text{q}_{b}\text{(5530)}}\text{q}_{C}%
\text{(2270)}$ & -1220 & B$_{C}$(6580) & 2.8\\\hline
$\Upsilon${\small (9460)} & $\overline{\text{q}_{b}\text{(5530)}}\text{q}%
_{b}\text{(5530)}$ & -1671 & $\Upsilon${\small (9389)} & 0.8\\\hline
\end{tabular}
\ \ $\\
\ \ \ \ {\small In the fifth column, R =(}$\frac{\Delta\text{M}}{\text{M}}%
${\small )\%.}%
\end{tabular}
$ \ \ \ \ 

\ 

The theoretical quantum numbers (I, S, C, B, Q) of the important mesons are
completely the same with the experimental results. The theoretical masses of
the important mesons agree\ well with the experimental results. These show
that the theoretical masses of the quarks\ are correct.\ 

\section{Predictions}

This proposal\ predicts some quarks and baryons.

\subsection{The
Quarks\ \ \ \ \ \ \ \ \ \ \ \ \ \ \ \ \ \ \ \ \ \ \ \ \ \ \ \ \ \ \ \ \ \ \ \ \ \ \ \ \ \ \ \ \ \ \ \ \ \ \ \ \ \ \ \ \ \ \ \ \ \ \ \ \ \ \ \ \ \ \ \ \ \ \ \ \ \ \ \ \ \ \ \ \ \ \ \ \ \ \ \ \ \ \ \ \ \ \ \ \ \ \ \ \ \ \ \ \ \ \ \ \ \ \ \ \ \ \ \ \ \ \ \ \ \ \ \ \ \ \ \ \ \ \ \ \ \ \ \ \ \ \ \ \ \ \ \ \ \ \ \ \ \ \ \ \ \ \ \ \ \ \ \ \ \ \ \ \ \ \ \ \ \ \ \ \ \ \ \ \ \ \ \ \ \ \ \ \ \ \ \ \ \ \ \ \ \ \ \ \ \ \ \ \ \ \ \ \ \ \ \ \ \ \ \ \ \ \ \ \ \ \ \ \ \ \ \ \ \ \ \ \ \ \ \ \ \ \ \ \ \ \ \ \ \ \ \ \ \ \ \ \ \ \ \ \ \ \ \ \ \ \ \ \ \ \ \ \ \ \ \ \ \ \ \ \ \ \ \ \ \ \ \ \ \ \ \ \ \ \ \ \qquad
\ \ \ \ \ \ \ \ \ \ \ \ \ \ \ \ \ \ \ \ \ \ \ \ \ \ \ \ \ \ \ \ \ \ \ \ \ \ \ \ \ \ \ \ \ \ \ \ \ \ \ \ \ \ \ \ \ \ \ \ \ \ \ \ \ \ \qquad
}

The new excited quarks: {\small d}$_{S}${\small (1390), } {\small u}$_{c}%
$(6490) and d$_{b}$(9950).

\subsection{The Baryons
\ \ \ \ \ \ \ \ \ \ \ \ \ \ \ \ \ \ \ \ \ \ \ \ \ \ \ \ \ \ \ \ \ \ \ \ \ \ \ \ }%

For these new quarks, there will be the new
baryons:\ \ \ \ \ \ \ \ \ \ \ \ \ \ \ \ \ \ \ \ \ \ \ \ \ \ \ \ \ \ \ \ \ \ \ \ \ \ \ \ \ \ \ \ \ \ \ \ \ \ \ \ \ \ \ \ \ \ \ \ \ \ \ \ \ \ \ \ \ \ \ \ \ \ \ \ \ \ \ \ \ \ \ \ \ \ \ \ \ \ \ \ \ \ \ \ \ \ \ \ \ \ \ \ \ \ \ \ \ \ \ \ \ \ \ \ \ \ \ \ \ \ \ \ \ \ \ \ \ \ \ \ \ \ \ \ \ \ \ \ \ \ \ \ \ \ \ \ \ \ \ \ \ \ \ \ \ \ \ \ \ \ \ \ \ \ \ \ \ \ \ \ \ \ \ \ \ \ \ \ \ \ \ \ \ \ \ \ \ \ \ \ \ \ \ \ \ \ \ \ \ \ \ \ \ \ \ \ \ \ \ \ \ \ \ \ \ \ \ \ \ \ \ \ \ \ \ \ \ \ \ \ \ \ \ \ \ \ \ \ \ \ \ \ \ \ \ \ \ \ \ \qquad
\ \ \ \ \ \ \ \ \ \ \ \ \ \ \ \ \ \ \ \ \ \ \ \ \ \ \ \ \ \ \ \ \ \ \ \ \ \ \ \ \ \ \ \ \ \ \ \ \ \ \ \ \ \ \ \ \ \ \ \ \ \ \ \ \ \ \ %

\begin{tabular}
[c]{ll}%
\lbrack d$_{S}^{0}$(1390)u'(3)d'(7)] =\ $\Lambda$(1400) & $\Lambda$(1405)\\
\lbrack u$_{C}^{0}$(6490)u'(3)d'(7)] =$\Lambda_{c}^{+}$(6500) &
\ \ \ \ \ \ ?\\
\lbrack d$_{b}^{0}$(9950)u'(3)d'(7)] = $\Lambda_{b}^{0}$(9960) & \ \ \ \ \ \ ?
\end{tabular}
\ \ \ \ \ \ \ \ \ \ \ \ \ \ \ \ \ \ 

The baryon $\Lambda$(1405) has already been discovered by experiments. The
baryons $\Lambda_{c}^{+}$(6500) and $\Lambda_{b}^{0}$(9960) are waiting to be discovered.

\section{Discussions\ \ \ \ \ \ \ \ \ \ \ \ \ \ \ \ \ \ \ \ \ \ \ \ \ \ \ \ \ \ \ \ \ \ \ \ }%

1. From (\ref{Alpha}) and (\ref{360 Mev}), we have
\begin{equation}
\text{m}_{q}\text{a}^{2}\text{ = h}^{2}\text{/720 Mev.}\label{Mba}%
\end{equation}
Although we do not know the values of $m_{b}$ and $a$, we find that
$m_{b}a^{2}$ is a constant. Since the Standard Model works fine without
considering the quark lattice, we estimate a $\leqq$ 10$^{-18}$m (the distance
scales limit of the standard model)\ \cite{Standard}. Thus the bare mass
(m$_{q}$) of the elementary quarks is
\begin{equation}%
\begin{tabular}
[c]{l}%
$\text{m}_{q}\text{{}}=\text{{}}\frac{\text{h}^{2}}{\text{\ \ 720\textit{a}%
}^{2}}\text{ }\geqq\text{ }\frac{\text{43.90}\times\text{ 10}^{-66}\text{(J
s)}^{2}}{\text{720}\times\text{ 10}^{6}\times\text{\ 1.602}\times\text{
10}^{-19}\text{J\ }\times\text{\ 10}^{-36}\text{m}^{2}}$\\
$\text{\ m}_{q}\text{ }\geqq\text{ 3.8 }\times\text{\ \ 10}^{-15}\text{kg =
2.27 }\times\text{ \ 10}^{12}\text{m}_{p}$\\
$\ \ \ \ \ \ \ \geqq$ 2.129 $\times\text{10}^{15}$Mev$.$%
\end{tabular}
\label{Bare M}%
\end{equation}
It is much larger than the excited quark rest masses. This ensures that the
Schr\"{o}dinger equation is a good approximation of the special quark Dirac
equation for deducing the rest quark masses and that the Standard Model is an
excellent approximation of the unborn fundamental theory at the distance
scales
$>$
10$^{-18}$m.

2. After the discovery of superconductors, we could understand the vacuum
material. In a sense, the vacuum material (skeleton-- the BCC quark lattice)
works like a superconductor. Since the transition temperature is very high
(much higher than the temperature at the center of the sun), all phenomena
that we can see are under the transition temperature. Thus\ there are no
electric or mechanical resistances to any particle or to any physical body
moving inside the vacuum material. As they move (with a constant velocity)
inside it, they look as if they are moving in completely empty space. The
vacuum material is a super superconductor.

3. There is an approximation (the quark Schr\"{o}dinger equation instead of
the low energy special quark Dirac equation) used in deducing the rest masses
of the quarks. The theoretical baryon spectrum (Table 2) and the theoretical
meson spectrum (Table 3) agree well with the experimental results. These
strong agreements mean that this approximation is good. Since the rest masses
are the same in any reference, in order to deduce the rest masses, we can use
the Schr\"{o}dinger equation instead of the special quark Dirac equation
\ \cite{Schrodinger} to deduce the approximate rest masses.

4. The flavored quarks (the flavored baryons and mesons) originated from the
body center cubic periodic symmetries of the BCC quark lattice and the
fluctuation. If BCC symmetries did not exist, there would not be any flavored
quark (baryon or meson).

5. The proposal has shown that the u-quark and the c-quark are the excited
states of the elementary u(0)-quark and that the d-quark, the s-quark and the
b-quark are the excited states of the elementary d(0)-quark. The u(0)-quark
and the d(0)-quark have a SU(2) symmetry (u(0) and d(0)). Therefore SU(3) (u,
d and s), SU(4) (u, d, s and c) and SU(5) (u, d, s, c and b) are correct
although there are large differences in masses between the quarks. In fact,
the SU(3), SU(4) and SU(5) are natural expansions of the SU(2). Since the bare
masses of the elementary quarks u(0) and d(0) (\ref{Bare M}) are huge
\begin{equation}
\text{\ m}_{\text{u(0)}}\text{(or m}_{\text{d(0)}})\text{ }\geqq\text{
2.27}\times\text{10}^{12}\text{m}_{p}=\text{2.129}\times\text{10}%
^{15}\text{Mev,}\label{BareM}%
\end{equation}
thus the bare masses (taking the absolutely empty spece as the zero energy
point) of the quarks (u, d, s, c, b, u' and d') are huge too:%

\begin{equation}%
\begin{tabular}
[c]{|l|l|}\hline
$\text{m}_{u}^{bare}\text{=\ m}_{u(0)}$+930 & $\text{m}_{d}^{bare}%
\text{=\ m}_{d(0)}$+930\\\hline
$\text{m}_{c}^{bare}\text{=\ m}_{u(0)}$+2270 & $\text{m}_{s}^{bare}%
\text{=\ m}_{d(0)}$+1110\\\hline
$\text{m}_{b}^{bare}\text{=\ m}_{d(0)}$+5530 & $\text{m}_{u\text{'}}%
^{bare}\text{=\ m}_{u(0)}$+ 3\\\hline
$\text{m}_{d}^{bare}\text{=\ m}_{d(0)}$+ 7 & \ \ .\\\hline
\end{tabular}
\label{Qb}%
\end{equation}
From (\ref{Bare M}) and (\ref{Qb}), we can see that the bare masses of the
quarks (u, d, s, c, b, u' and d') are essentially the same. This is a rigorous
physical basis of the SU(3), SU(4), SU(5) and so forth symmetries. We had been
thinking that SU(4) and SU(5) do not have rigorous physical basis for a long
time since the differences of the quark masses are too large. Considering the
bare masses of the quarks, now we believe that SU(4) and SU(5) symmetries
really exist.

6. After we study the BCC quark lattice (with a periodic constant a $\leq$
10$^{-18}$m)\ and deduce the masses and the intrinsic quantum numbers (I, s,
c, b and Q) of the quarks, we have a deeper understanding of the Standard
Model. When the distance scales
$>$
10$^{-18}m$, although we cannot see the quark lattice, we will see the
u(0)-quark Dirac sea and the d(0)-quark Dirac sea. Sometimes, we can also see
the s-quark, the c-quarkand the b-quark (inside baryons and mesons); and, from
the Dirac sea concept, we guess that there will be an s-quark Dirac sea, a
c-quark Dirac sea and a d-quark Dirac sea too. Since we cannot see the quark
lattice (we can only see the Dirac seas), we cannot deduce the masses and the
intrinsic quantum numbers (experimental measurements). Naturally we think that
the quarks (u, d, s, c and b) are all independent elementary particles. The
Standard Model is a reasonably excellent approximation to nature at distance
scales as small as 10$^{-18}$m \cite{Standard}. Thus there is no contradiction
between the Standard Model and the BCC quark lattice Model, otherwise the BCC
quark lattice model does provide a physical basis for the Standard Model.

\section{Conclusions\ \ \ \ \ \ \ \ \ \ \ \ \ \ \ \ \ \ \ \ \ \ \ \ \ \ \ \ \ \ \ \ \ \ }%

1.\ The origin of the quark masses is the strong interactions between the BCC
quark lattice and the excited quarks. Using the strong interactions, we have
deduced that m(u)=930 Mev, m(d)=930 Mev, m(s)=1110 Mev, m(c)=2270 Mev and
m(b)=5530 Mev. The theoretical masses (deduced from the quark masses) of the
most important baryons (Table 2) and mesons (Table 3) agree well with the
experimental results. This shows that the theoretical masses of the quarks are correct.

2. There are only two elementary quarks (u(0) and d(0)) in the vacuum state.
All quarks (u, d, s, c, b ...) with rest mass $\neq0$ are the energy band
excited states of u(0) and d(0).\ The quarks u, d, s, c and b are the ground
states of them. The quarks u(930) and u$_{C}$(2270) are the excited states of
the elementary u(0)-quark: the quarks d(930), s(1110) and b(5530) are the
excited states of the elementary d(0)-quark.

3. The SU(3), SU(4) and SU(5) are natural expansions of the SU(2).

4. The BCC quark lattice model provides a rigorous physical basis for the
Quark Model.

5. Due to the existence of the vacuum quark lattice, all observable particles
are constantly affected by the lattice. Thus some laws of statistics (such as
fluctuation) cannot be ignored.

\begin{center}
\bigskip\textbf{Acknowledgment}
\end{center}

I would like to express my heartfelt gratitude to Dr. Xin Yu for checking the
calculations of the energy bands. I sincerely thank Professor Robert L.
Anderson for his valuable advice. I acknowledge\textbf{\ }my indebtedness to
Professor D. P. Landau for his help also.

\section{Appendix: The Body Center Cubic Quark Lattice}

According to Dirac's sea concept \cite{DiracSea}, there is an electron-Dirac
sea, a $\mu$-lepton Dirac sea, a $\tau$-lepton Dirac sea, a $u$-quark Dirac
sea, a $d$-quark Dirac sea, an $s$-quark Dirac sea, a $c$-quark Dirac sea, a
$b$-quark Dirac sea and so forth in the vacuum. All of these Dirac seas are in
the same space, at any location. These particles will interact with one
another and form the perfect physical vacuum material. Some kinds of
particles, however, do not play an important role in forming the physical
vacuum material. First, the main force which makes and holds the structure of
the physical vacuum material must be the strong interactions, not the
weak-electromagnetic interactions. Hence, in considering the structure of the
vacuum material, we leave out the Dirac seas of those particles which do not
have strong interactions ($e$, $\mu$ and $\tau$). Secondly, the physical
vacuum material is super stable, hence we also omit the Dirac seas which can
only make unstable baryons (the $s$-quark, the $c$-quark and the $b $-quark).
Finally, there are only two kinds of possible particles left: the vacuum state
$u(0)$-quarks and the vacuum state $d(0)$-quarks. There are super strong
attractive forces between the $u(0)$-quarks and the $d(0)$-quarks (colors)
that will make and hold the densest structure of the vacuum material.

According to solid state physics \cite{BodyCenter}, if two kinds of particles
(with radius $R_{1}<R_{2}$) satisfy the condition $1>R_{1}/R_{2}>0.73$, the
densest structure is the body center cubic lattice \cite{BCC}. We know the
following: first, $u$ quarks and $d$ quarks are not exactly the same, thus
$R_{u}\neq R_{d}$; second, they are very close to each other (with the same
bare masses essentially), thus $R_{u}\approx R_{d}$. Hence, if $R_{u}<R_{d}$
(or $R_{d}<R_{u}$), we have $1>R_{u}/R_{d}>0.73$ (or $1>R_{d}/R_{u}>0.73$).
Therefore, we conjecture that the vacuum state u(0)-quarks and d(0)-quarks
construct the body center cubic quark lattice in the vacuum.

\end{document}